# 13 New Light Curves and Updated Mid-Transit Time and Period for Hot Jupiter WASP-104 b with EXOTIC

**Heather B. Hewitt**
*School of Earth and Space Exploration, Arizona State University, 781 E Terrace Mall, Tempe, AZ 85287-6004; hbhewittt@asu.edu*

| | | |
|---|---|---|
| **Federico Noguer** | **Kendall Collins** | **Chyna Merchant** |
| **Suber Corley** | **Kris Ganzel** | **Matthew Pedone** |
| **James Ball** | **Kimberly Merriam Gray** | **Gina Plumey** |
| **Claudia Chastain** | **Mike Logan** | **Matthew Rice** |
| **Richard Cochran-White** | **Steve Marquez-Perez** | **Zachary Ruybal** |

*School of Earth and Space Exploration, Arizona State University, 781 E Terrace Mall, Tempe, AZ 85287-6004*

**Molly N. Simon**
*School of Earth and Space Exploration, Arizona State University, 781 E Terrace Mall, Tempe, AZ 85287-6004*

**Isabela Huckabee**
*School of Earth and Space Exploration, Arizona State University, 781 E Terrace Mall, Tempe, AZ 85287-6004 and Jet Propulsion Laboratory, California Institute of Technology, 4800 Oak Grove Drive, Pasadena, CA 91109*

**Robert T. Zellem**
*Jet Propulsion Laboratory, California Institute of Technology, 4800 Oak Grove Drive, Pasadena, CA 91109*

**Kyle A. Pearson**
*Jet Propulsion Laboratory, California Institute of Technology, 4800 Oak Grove Drive, Pasadena, CA 91109*

**Abstract**  Using the EXOplanet Transit Interpretation Code (EXOTIC), we reduced 52 sets of images of WASP-104 b, a Hot Jupiter-class exoplanet orbiting WASP-104, in order to obtain an updated mid-transit time (ephemeris) and orbital period for the planet. We performed this reduction on images taken with a 6-inch telescope of the Center for Astrophysics | Harvard & Smithsonian MicroObservatory. Of the reduced light curves, 13 were of sufficient accuracy to be used in updating the ephemerides for WASP-104 b, meeting or exceeding the three-sigma standard for determining a significant detection. Our final mid-transit value was $2457805.170208 \pm 0.000036$ BJD_TBD and the final period value was $1.75540644 \pm 0.00000016$ days. The true significance of our results is in their derivation from image sets gathered over time by a small, ground-based telescope as part of the Exoplanet Watch citizen science initiative, and their competitive results to an ephemeris generated from data gathered by the TESS telescope. We use these results to further show how such techniques can be employed by amateur astronomers and citizen scientists to maximize the efficacy of larger telescopes by reducing the use of expensive observation time. The work done in the paper was accomplished as part of the first fully online Course-Based Undergraduate Research Experience (CURE) for astronomy majors in the only online Bachelor of Science program in Astronomical and Planetary Sciences.

## 1. Introduction

The study of exoplanets is a popular and fast-growing subject in astronomy. By studying the variety of extrasolar planets and planetary systems that have been discovered, researchers gain a deeper understanding of how planetary formation and evolution occur and gain valuable insights into the composition of other distant worlds. Currently, over 5,000 exoplanets have been confirmed, up from 32 in 2000. Historically, opportunities for research on exoplanets have been limited for amateur astronomers and those without professional backgrounds in astronomy. High costs and technical expertise are some of the obstacles to building, operating, or maintaining appropriate observational equipment. However, citizen science projects, including Exoplanet Watch, aim to expand the usefulness of direct observations of transiting exoplanets with a network of small Earth-based telescopes (Zellem *et al.* 2020). Similarly, Exoplanet Watch and others increase the efficiency of exoplanet studies conducted by large telescopes by reducing uncertainty about the predicted timing of transit events (Zellem *et al.* 2020).

Improving the potentially stale ephemerides of known exoplanet transits is now an established method of reducing observational costs for space telescopes (e.g. Zellem *et al.* 2020; Kokori *et al.* 2022b; Yeung *et al.* 2022). The work done in this paper contributes to the network of small telescope observations that funnel improved ephemerides to established repositories for use by scientists conducting large space- and ground-based telescope observation missions. Given the large cost of using space-based telescopes, these improvements represent thousands of dollars in cost savings by improving efficiency (Drier 2021). The work that amateur astronomers and citizen scientists do when working with teams such as Exoplanet Watch ensures that future use of expensive time on telescopes is used efficiently. This project partnered with Exoplanet Watch to examine previously unreviewed astronomical data for the planet WASP-104 b. WASP-104 b is a characteristic Hot Jupiter with a mass of about 1.272 Jupiter masses, a period of about 1.75 days, and orbits at a distance of 0.029 AU from its host star, WASP-104 (Smith *et al.* 2014). WASP-104 is a G-type star.

The work done in the paper was accomplished as part of the first fully online Course-Based Undergraduate Research Experience (CURE) for astronomy majors. Fifteen students participated in the 15-week online course, Exoplanet Research Experience, at Arizona State University (ASU). This course was developed to enhance the only completely online Bachelor of Science program in Astronomical and Planetary Sciences (APS). The APS degree program was developed to mirror the existing in-person Astrophysics degree program at ASU, but at the time this course was developed, there was no opportunity for the online students to participate in authentic research experiences. This is a common disparity between online and in-person degree programs that we aimed to address with the development of this CURE.

**2. Observatory**

We obtained our data from the MicroObservatory Robotic Telescope Network, which is operated by the Harvard Smithsonian Center for Astrophysics (Sadler *et al.* 2001). The MicroObservatory uses a network of robotic 6-inch telescopes. Our observations were taken using Cecilia. Cecilia is part of the MicroObservatory network and is located on Mount Hopkins, Arizona, at the Whipple Observatory. It is a custom-built Maksutov-Newtonian with an aperture of 152 mm and a focal length of 560 mm. It is equipped with a KAF-1402ME camera and produces $0.94 \times 0.72$ degree images; the images are binned $2 \times 2$.

**3. Weather**

Observations by ground-based telescopes have two main environmental factors with which they have to contend to produce favorable data: atmospheric turbulence and weather phenomena. Any combination of environmental and technical issues can hamper ground-based observations, which is a hindrance when compared to orbital telescopes. However, a ground-based telescope can make up for this deficit by the sheer volume of observations. Table 1 includes the date of each observation, and the average quality of the weather as estimated by Cecilia in the FITS header. The bolded dates in Table 1 indicate a significant detection. The weather is rated on a scale of 0 to 100. A 0 score represents a completely cloudy night, whereas a score of 100 represents a completely clear night. It's important to note that the MicroObservatory weather ratings were not available between 2018 and 2020; therefore, the weather ratings listed in Table 1 from 2018-02-23 to 2020-01-13 are not accurate (Sienkiewicz 2022). Additionally, we were able to remove some images with significant cloud cover from the data sets with low weather rankings to still obtain a significant detection in some cases. These weather ratings are used as a guide; they are estimated from NOAA weather satellites and do not always accurately reflect the local weather conditions. A further analysis of the data is required to determine the quality for each night of data.

Table 1. Weather quality estimates by average WEATHER value in Cecilia data: (Bold lines indicate a significant detection).

| Date | Weather Quality | Date | Weather Quality |
|---|---|---|---|
| **2015-02-07** | **98.74** | 2017-03-09 | 99.01 |
| 2015-02-14 | 39.09 | 2017-03-31 | 98.95 |
| 2015-02-21 | 22.83 | **2017-05-13** | **99.78** |
| 2015-02-22 | 14.43 | 2018-01-04 | 0.00 |
| **2015-02-28** | **84.10** | 2018-02-23 | 4.00 |
| 2015-03-22 | 91.93 | 2018-03-03 | 4.00 |
| **2015-04-05** | **79.37** | 2018-03-10 | 4.00 |
| **2015-05-12** | **38.64** | 2018-03-17 | 4.00 |
| **2015-05-26** | **99.27** | 2018-03-18 | 4.00 |
| 2015-06-02 | 99.82 | 2018-03-24 | 4.00 |
| 2016-01-03 | 73.88 | 2018-03-25 | 4.00 |
| 2016-01-10 | 36.47 | **2018-04-01** | **4.00** |
| 2016-01-17 | 10.22 | 2018-04-22 | 4.00 |
| 2016-02-22 | 98.66 | 2018-05-06 | 4.00 |
| **2016-02-29** | **98.33** | 2018-11-23 | 4.00 |
| 2016-03-07 | 3.54 | 2018-11-29 | 4.00 |
| 2016-03-23 | 98.68 | 2019-01-19 | 4.00 |
| 2016-04-06 | 0.20 | 2019-02-17 | 4.00 |
| **2016-04-13** | **98.83** | 2019-02-24 | 4.00 |
| **2016-04-21** | **99.00** | 2019-03-18 | 4.00 |
| 2016-11-28 | 55.62 | 2020-01-06 | 4.00 |
| 2016-12-05 | 87.03 | **2020-01-13** | **4.00** |
| 2017-01-17 | 68.51 | 2020-01-28 | 96.72 |
| **2017-02-22** | **98.38** | 2020-02-25 | 96.53 |
| 2017-02-23 | 98.55 | 2020-03-11 | 0.00 |
| 2017-03-01 | 96.25 | 2020-03-18 | 59.76 |

**4. Data reduction**

Our team utilized NASA Jet Propulsion Laboratory's software, EXOTIC (EXOplanet Transit Interpretation Code), to analyze our photometric data and reduce the light curves for our 52 nights of data (Zellem *et al.* 2020). EXOTIC reduces raw ".fits" files into a light curve and calculates target parameters by tracking the target throughout the observation. EXOTIC is a Python 3 pipeline that can be run locally or on Google Colab. We chose to run EXOTIC in the Google Colab Cloud, which supports the sharing of files among team members. In the Google Colab, we mounted our data and installed EXOTIC onto a virtual machine. Priors for the target are obtained from the NASA Exoplanet Archive by searching the target's name. Then, an image is displayed, and users are prompted to locate the target and up to ten comparison stars. To determine the flux of the target, an optimal aperture size is determined and all the pixel values within the aperture are summed. The background light is subtracted from each pixel value to isolate the flux from the star itself. To ensure the star's brightness is changing due to a transit and not to atmospheric interference, EXOTIC compares the star's brightness to the brightness of nearby comparison stars (Zellem *et al.* 2020). EXOTIC's output includes a light curve, the mid-transit time, ratio of planet to stellar radius, transit depth, ratio of semi-major axis to stellar radius, scatter in residuals, and transit duration.

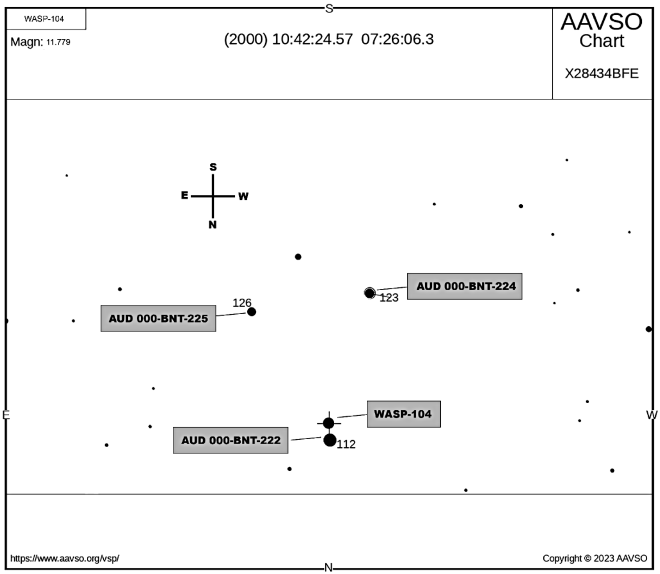

Figure 1. AAVSO VSP Chart for WASP-104 b with comparison stars labeled. Original field of view of AAVSO VSP Chart was 18.5 arc minutes. The image was magnified for easier viewing.

Table 2. Calculated mid-transit times and transit depths for WASP-104 b.

| Date | Transit Depth (%) | Transit Depth Uncertainty | Mid-transit (BJD_TBD) | Mid-transit Uncertainty (d) |
|---|---|---|---|---|
| 2015-02-07 | 2.30 | 0.51 | 2457060.8809 | 0.0026 |
| 2015-02-28 | 2.29 | 0.44 | 2457081.9381 | 0.0028 |
| 2015-04-05 | 2.30 | 0.3  | 2457118.8094 | 0.0025 |
| 2015-05-12 | 2.29 | 0.53 | 2457155.6699 | 0.0032 |
| 2015-05-26 | 2.30 | 0.16 | 2457169.7162 | 0.0029 |
| 2016-03-01 | 2.26 | 0.68 | 2457448.8305 | 0.0042 |
| 2016-04-14 | 2.30 | 0.41 | 2457492.7111 | 0.0032 |
| 2016-04-21 | 2.30 | 0.45 | 2457499.7262 | 0.0041 |
| 2017-02-22 | 2.30 | 0.18 | 2457806.9359 | 0.0023 |
| 2017-05-13 | 2.30 | 0.47 | 2457887.6848 | 0.0035 |
| 2018-04-01 | 2.30 | 0.16 | 2458210.6607 | 0.0039 |
| 2020-01-13 | 2.27 | 0.27 | 2458861.9278 | 0.0024 |
| 2020-02-25 | 2.06 | 0.22 | 2458905.8001 | 0.0035 |

In order to confirm which data sets we intended to use, our research team used an agreed-upon control method to ensure that a uniform process was followed before we finalized our list of significant detections. This agreed-upon method of analyzing the data included defining which images are candidates for deletion and identifying the best comparison stars to use for every data set. First, we cleaned the images up by removing "bad" images, defined as having clouds, an unacceptable level of visual noise, or the target missing from the field of view. This assessment was conducted using astronomical image-viewing programs SAOImage DS9, AstroImageJ, or JS9. Then, using the American Association of Variable Star Observer (AAVSO) Variable Star Plotter (VSP; AAVSO 2023), we selected three comparison stars to use in our reduction process:

AUD 000-BNT-222, AUD 000-BNT-224, and AUD 000-BNT-225. Our chart (Chart ID X27938O) is shown in Figure 1 with our comparison stars labeled.

With our agreed-upon selection process, we re-analyzed the 16 possibly significant detections of the original 52 observations and produced 13 confirmed significant detections. We determined a detection as significant when the detection significance was greater than 3σ using Equation (1):

$$\frac{\text{Transit Depth}}{\text{Transit Depth Uncertainty}} \geq 3 \quad (1)$$

### 5. Data

From 52 nights worth of data and images, we reduced 13 significant light curves using EXOTIC's reduction process on the target hot Jupiter, WASP-104 b, as shown in Table 2. A sample light curve is shown in Figure 2 and all the light curves are presented in Appendix A, Figure A1.

We created an observed-calculated (O–C) plot to calculate our updated mid-transit time (ephemeris) and period. We also used a posterior distribution to analyze our parameters statistically. A secondary tool in Colab[1] was used that fit the ephemeris of our observations to previously published observations in the Exoplanet Archive. In the O–C plot, shown in Figure 3, we included the mid-transit times from our 13 significant detections along with two of the mid-transit times on the NASA Exoplanet Archive: Smith *et al.* (2014) and the most recent, Ivshina and Winn (2022). For consistency, we chose only to include previously published mid-transit times derived from a measured light curve. As such, we excluded the Bonomo *et al.* (2017) and Kokori *et al.* (2022a) values from the O–C plot. The values used from the NASA Exoplanet Archive are shown in Table 3. We used the most recently published mid-transit time (2457805.170205 ± 0.000037) and period (1.75540569 ± 0.00000011) as our priors (Ivshina and Winn 2022). The ephemeris fitter calculated our updated mid-transit time to be 2457805.170208 ± 0.000036 BJD_TDB and our updated period to be 1.75540644 ± 0.00000016 days. The posterior plot distribution for our new mid-transit time and period are shown in Figure 4.

### 6. Results

Over time, the uncertainties of mid-transit times become stale, so, in order to accurately compare our mid-transit uncertainties to those published previously, it was necessary to forward-propagate the previously published mid-transit times to our newly updated mid-transit time. In order to do this, we used Equation (2) from Zellem *et al.* (2020):

$$\Delta T_{mid} = (n_{orbit}^2 \cdot \Delta P^2 + 2n_{orbit} \cdot \Delta P \Delta T_0 + \Delta T_0^2)^{1/2} \quad (2)$$

Following Zellem *et al.* (2020), we dropped the second term of Equation 2 because none of the previous publications report their covariance term. This leads the propagated mid-transit uncertainties to be slightly underestimated. After forward-propagating the mid-transit times published by Smith et al. (2014), Kokori et al. (2022a), and Ivshina and Winn (2022), we found the new times to be 2456406.11126 ± 0.0014, 2457048.59061 ± 0.00021, and 2457805.170205 ± 0.000037, respectively. The Ivshina and Winn (2022) mid-transit uncertainty remains unchanged due to how recently it was calculated. Our updated mid-transit time, mid-transit uncertainty, period, and period uncertainty for WASP-104 b are presented in Table 4 along with the original mid-transit times and mid-transit uncertainty values from Smith et al. (2014), Kokori et al. (2022a), and Ivshina and Winn (2022) and their respective propagated mid-transit uncertainties. Comparing these propagated mid-transit times to our updated mid-transit time, we have decreased the mid-transit uncertainty by 97.4% since the discovery paper (Smith et al. 2014). We also decreased the Kokori et al. (2022) mid-transit uncertainty by 82.9% and nearly matched the results from Ivshina and Winn (2022), slightly decreasing the mid-transit uncertainty by 2.7%. We also compared the uncertainty in our reported period to that reported by Smith et al. (2014), Kokori et al. (2022a), and Ivshina and Winn (2022). We decreased the period uncertainty from Smith et al. (2014) by 91% in the positive direction and 95.6% in the negative direction, decreased the period uncertainty from Kokori et al. (2022a) by 46.7%, and produced a slightly increased period uncertainty from Ivshina and Winn (2022) (by 45.5%).

---
[1] The ephemeris fitter can be found at https://colab.research.google.com/drive/1T5VT2gZ-ip6K6T9IXqMzQdSiEaf-UbJn?usp=sharing

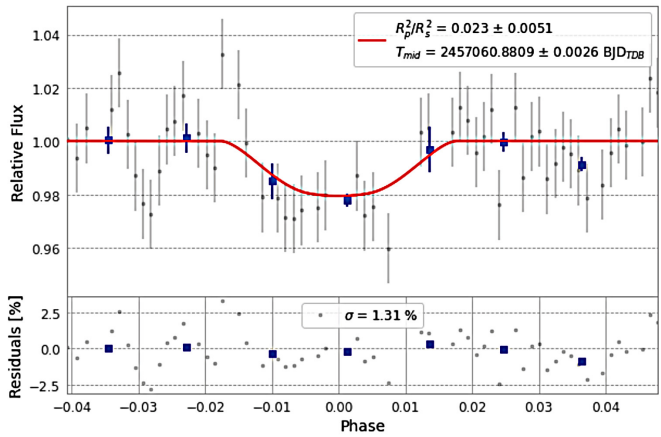

Figure 2. Sample light curve that was reduced from data taken on 2015-02-07. The gray points represent data from each image in the data set. The blue points represent the average of a set of binned data points, used to fit the light curve.

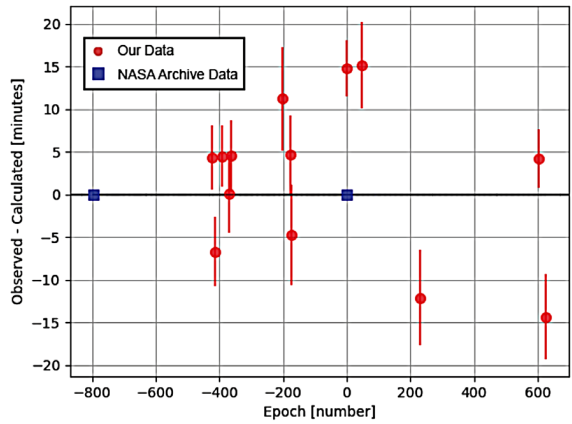

Figure 3. O–C plot for WASP-104 b using $t_0$ = 2457805.170205 BJD_TDB and P = 1.75540569 days.

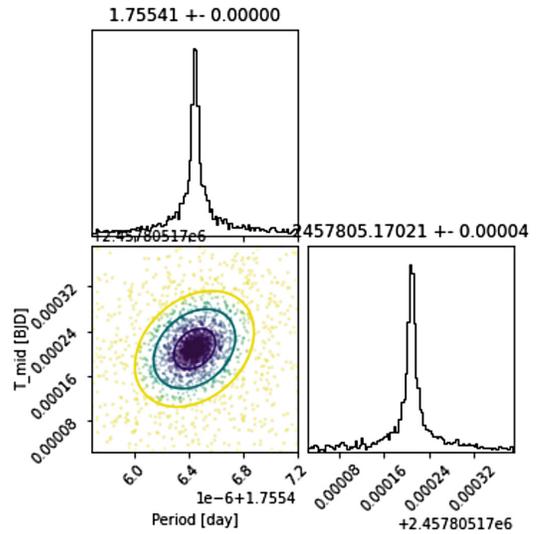

Figure 4. Posterior plot distribution of our new mid-transit time and period. The data points are color-coded to the likelihood of each fit, with darker colors indicating a higher likelihood.

Table 3. Values from the NASA Exoplanet Archive used in the creation of the O–C plot.

| Citation | Mid-transit (BJD_TBD) | Mid-transit Uncertainty (d) |
| --- | --- | --- |
| Smith et al. (2014) | 2456406.11126 | 0.00012 |
| Ivshina and Winn (2022) | 2457805.170205 | 0.000037 |

Table 4. Updated Ephemerides.

| Citation | Mid-transit (BJD_TBD) | Mid-transit Uncertainty (d) | Propagated Mid-transit Uncertainty (d) | Period (d) | Period Uncertainty (d) |
| --- | --- | --- | --- | --- | --- |
| Our data | 2457805.170208 | 0.000036 | N/A | 1.75540644 | 0.00000016 |
| Smith et al. (2014) | 2456406.11126 | 0.00012 | 0.0014 | 1.7554137 | +0.0000018 / −0.0000036 |
| Kokori et al. (2022a) | 2457048.59061 | 0.00016 | 0.00021 | 1.7554060 | 0.0000003 |
| Ivshina and Winn (2022) | 2457805.170205 | 0.000037 | 0.000037 | 1.75540569 | 0.00000011 |

Out of 52 data sets from Cecilia, we were able to reduce 13 light curves with an accuracy meeting or exceeding the three-sigma standard for a successful detection. In so doing, we managed to calculate mid-transit and period values for WASP-104 b with improved precision over previous results on the NASA Exoplanet Archive and rivaling those more recent results that relied on data from the orbital TESS telescope. We calculated our updated mid-transit time to be 2457805.170208 ± 0.000036 BJD_TDB and our updated period to be 1.75540644 ± 0.00000016 days.

The results achieved in the work done by the first-ever fully online research experience course that is documented here substantiate that the Exoplanet Watch model of using small-format terrestrial telescopes to gather observations and to process them through the EXOTIC pipeline is a cost-effective alternative to observations with large terrestrial and space telescopes. The changes in the ephemerides of WASP-104 b over the past few years illustrate the need to regularly refresh them, and the use of a low-cost solution for the required observations, as shown here, is a logical path to achieving this goal.

In addition to the value of the data, the results of this study imply the value for small ground-based telescope photometry by citizen and amateur astronomers. Within a short amount of time, amateur astronomers around the world are able to reduce light curves, analyze the data, and present findings that are a benefit to the science community. The updating of ephemerides is a key part of future observations by large telescopes. Space-based telescopes like the NASA Transiting Exoplanet Survey Satellite (TESS) can update the ephemerides of many planets with higher precision. With the number of current and future scientists studying exoplanets on the rise, we need a reliable process to accomplish these updates. Our method can be continued regardless of current space missions, budget, and availability of observing time.

The network of amateur astronomers participating with Exoplanet Watch and using EXOTIC have grown substantially, and it is a powerful tool for the study of extrasolar planets. Furthermore, Exoplanet Watch's network has the ability to be fluid and quickly coordinate observations and data reduction without long waits during the process of applying for a space-based or a large ground-based telescope's time. There are planets whose transit time is longer than observing time from any location. With the coordination of Exoplanet Watch's network around the world, an entire transit can be observed with such planets as shown (Zellem *et al.* 2020).

### 7. Conclusion

With 52 nights of observation from the MicroObservatory, collected with a 6-inch ground-based robotic telescope and using the reduction tool EXOTIC, we present 13 significant light curves as well as an updated mid-transit time and period for WASP-104 b. We were able to decrease the uncertainties in the mid-transit time and period compared to those published previously this year (Kokori *et al.* 2022) as well as achieve nearly identical mid-transit time and period uncertainties as those obtained using data from TESS (Ivshina and Winn 2022). The comparison of our updated ephemerides to recently published results demonstrates the importance of citizen science groups like Exoplanet Watch and the capabilities of small ground-based telescopes.

The work done in this paper was performed in the first fully online CURE for astronomy majors. Online course and degree programs make higher education accessible to a more diverse learner population (e.g. women, veterans, parents, persons with disabilities, students with full-time jobs, and students of color). The success of the first offering of Arizona State University's Exoplanet Research Experience demonstrates the importance of undergraduate research experiences. The educational benefits and affective outcomes of participation in this online CURE will be addressed in a future paper.

This project validated several paradigms in exoplanet astronomy and astronomy education and, in the process, confirmed the conjunction of these paradigms. This included (1) the use of the small terrestrial telescopes in the Exoplanet Watch network, (2) the use of EXOTIC as a reduction pipeline for exoplanet transit data, and (3) a large-scale collaborative approach to learning the concepts and tools that are used in the identification and ephemerides refreshment of exoplanets.

### 8. Acknowledgements

These observations were conducted with MicroObservatory, a robotic telescope network maintained and operated as an educational service by the Center for Astrophysics | Harvard & Smithsonian. MicroObservatory is supported by NASA's Universe of Learning under NASA award number NNX16AC65A to the Space Telescope Science Institute.

This publication makes use of data products from Exoplanet Watch, a citizen science project managed by NASA's Jet Propulsion Laboratory on behalf of NASA's Universe of Learning. This work is supported by NASA under award number NNX16AC65A to the Space Telescope Science Institute, in partnership with Caltech/IPAC, Center for Astrophysics|Harvard & Smithsonian, and NASA Jet Propulsion Laboratory.

This research has made use of the NASA Exoplanet Archive, which is operated by the California Institute of Technology, under contract with the National Aeronautics and Space Administration under the Exoplanet Exploration Program.

This work is supported by the National Science Foundation (NSF) under grant #IUSE 2121225

Part of the research was carried out at the Jet Propulsion Laboratory, California Institute of Technology, under contract with the National Aeronautics and Space Administration. Copyright 2023. All rights reserved.

### References

AAVSO. 2023, Variable Star Plotter
   (VSP; https://www.aavso.org/apps/vsp/).
Bonomo, A. S., *et al.* 2017, *Astron. Astrophys.*, **602A**, 107.
Drier, C. 2021, Planetary.org
   (https://www.planetary.org/articles/cost-of-the-jwst).
Ivshina, E. S., and Winn, J. N. 2022, *Astrophys. J., Suppl. Ser.*, **259**, 62.

Kokori, A., Tsiaras, A., Edwards, B., *et al.* 2022a, *Astrophys. J., Suppl. Ser.*, **258**, 40.
Kokori, A., *et al.* 2022b, *Exp. Astron.,* **53**, 547.
Sadler, P. M., *et al.* 2001, *J. Sci. Education Technol.*, **10**, 39.
Sienkiewicz, F. 2022, private communication (August 6, 2022).
Smith, A. M., *et al.* 2014, *Astron. Astrophys.*, **570A**, 64.
Yeung, P., Perian, Q., Robertson, P., Fitzgerald, M., Fowler, M., and Sienkiewicz, F. 2022. *J. Korean Astron. Soc.*, **55**, 111.
Zellem, R., *et al.* 2020, *Publ. Astron. Soc. Pacific*, **132**, 054401.

## Appendix A: Significant detections of WASP-104 b

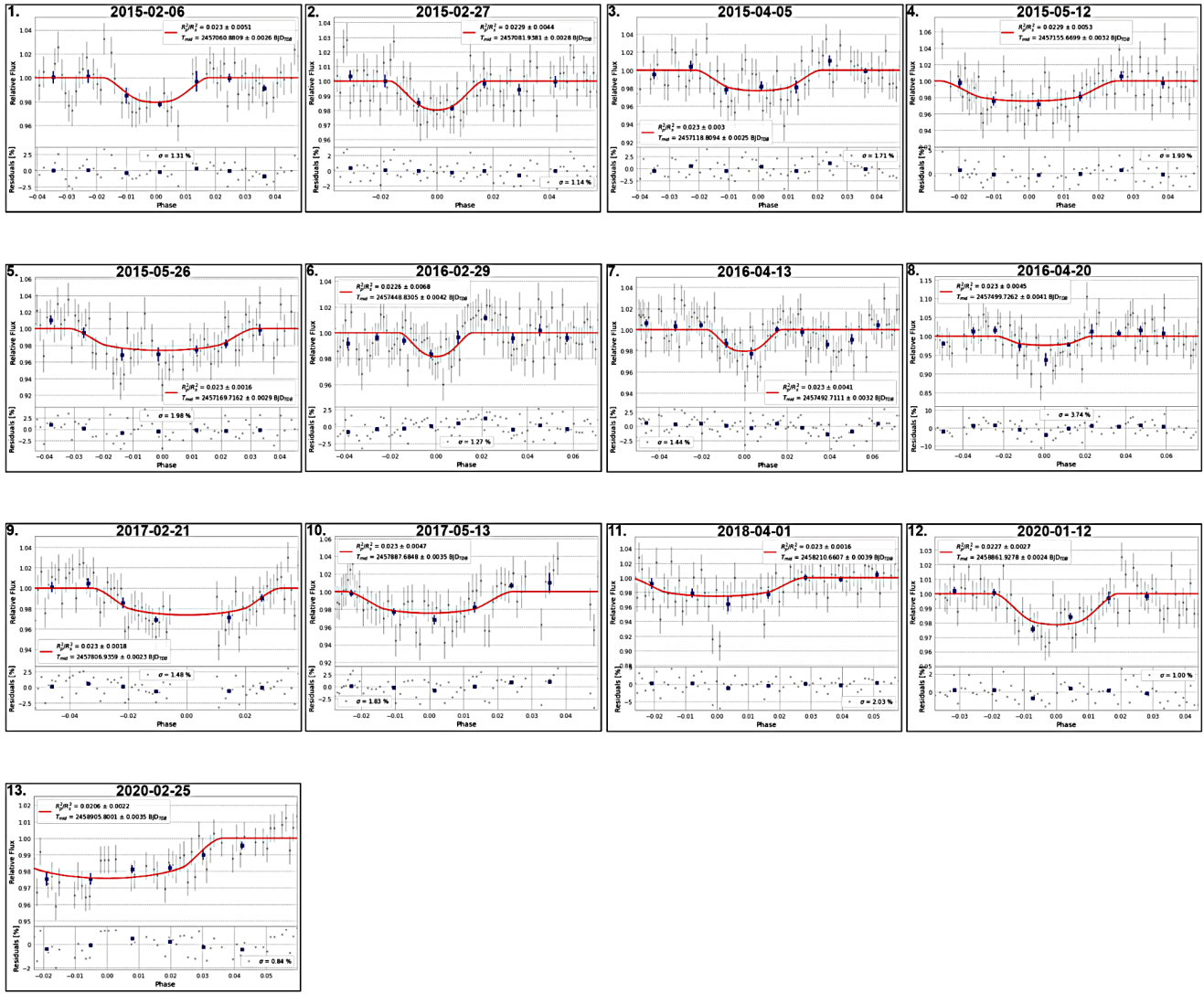

Figure A1. Light curves from this study.